\documentclass[journal,final,10pt]{IEEEtran}
\usepackage{pdfpages, comment}
\usepackage{epstopdf}
\usepackage{commath, amsmath, graphics, amssymb, epsfig, float, babel}
\usepackage{amsfonts, graphicx, color, upgreek}
\usepackage{cite, array, subfigure, multirow, hhline}
\usepackage{algorithm}
\usepackage{comment}
\usepackage[noend]{algpseudocode}

\def\BibTeX{{\rm B\kern-.05em{\sc i\kern-.025em b}\kern-.08em
    T\kern-.1667em\lower.7ex\hbox{E}\kern-.125emX}}
\def\FigureWidth{3.2in}

\usepackage[font=small]{caption} 

\newcommand{\be}{\begin{equation}}

\newcommand{\ee}{\end{equation}}
\newcommand{\bea}{\begin{eqnarray}}
\newcommand{\eea}{\end{eqnarray}}
\newcommand{\bdp}{\begin{displaymath}}
\newcommand{\edp}{\end{displaymath}}
\newcommand\NoDo{\renewcommand\algorithmicdo{}}

\floatstyle{ruled}
\newfloat{algorithm}{tbp}{loa}
\providecommand{\algorithmname}{Algorithm}
\floatname{algorithm}{\protect\algorithmname}
\usepackage[font=small]{caption} 

\makeatletter
\def\hlinewd#1{\noalign{\ifnum0=`}\fi\hrule \@height #1 \futurelet \reserved@a\@xhline}

\IEEEoverridecommandlockouts
\IEEEpubid{\makebox[\columnwidth]{ 978-1-6654-3540-6/22~\copyright~2022 IEEE \hfill} \hspace{\columnsep}\makebox[\columnwidth]{ }}
\begin{document}
\title{
Cooperative Multi-Agent Deep Reinforcement Learning Methods for UAV-aided Mobile Edge Computing Networks}

\author{\IEEEauthorblockN{Mintae Kim, Hoon Lee,~\IEEEmembership{Member,~IEEE}, Sangwon Hwang, Mérouane Debbah,~\IEEEmembership{Fellow,~IEEE} and Inkyu Lee,~\IEEEmembership{Fellow,~IEEE}}

\thanks{

M. Kim, S. Hwang, and I. Lee are with the School of Electrical Engineering, Korea University, Seoul 02841, Korea (e-mail: \{wkd2749, tkddnjs3510, inkyu\}@korea.ac.kr).

H. Lee is with the Department of Electrical Engineering and the AI graduate school, Ulsan National Institute of Science and Technology (UNIST), Ulsan, 44919, Korea (e-mail: hoonlee@unist.ac.kr).

M. Debbah is with Khalifa University of Science and Technology, P O Box 127788, Abu Dhabi, UAE (email: merouane.debbah@ku.ac.ae).}
} \maketitle
\begin{abstract}
This paper presents a cooperative multi-agent deep reinforcement learning (MADRL) approach for unmmaned aerial vehicle (UAV)-aided mobile edge computing (MEC) networks. An UAV with computing capability can provide task offlaoding services to ground internet-of-things devices (IDs). With partial observation of the entire network state, the UAV and the IDs individually determine their MEC strategies, i.e., UAV trajectory, resource allocation, and task offloading policy. This requires joint optimization of decision-making process and coordination strategies among the UAV and the IDs. To address this difficulty, the proposed cooperative MADRL approach computes two types of action variables, namely message action and solution action, each of which is generated by dedicated actor neural networks (NNs). As a result, each agent can automatically encapsulate its coordination messages to enhance the MEC performance in the decentralized manner. The proposed actor structure is designed based on graph attention networks such that operations are possible regardless of the number of IDs. A scalable training algorithm is also proposed to train a group of NNs for arbitrary network configurations. Numerical results demonstrate the superiority of the proposed cooperative MADRL approach over conventional methods.

\end{abstract}

\section{Introduction}
Mobile edge computing (MEC) systems have been regarded as promising solutions to provide remote computation services for internet-of-things (IoT) networks with the aid of edge servers \cite{NAbbas:IoTJ18,CChen_MEC:TMC23,LZhang_MEC:TC21,MMasoudi_MEC:TMC21,KZhang:CM18, MTKim:24, JHPark:TVT21, MWu:TVT21, NKiran:JCN23}. Edge servers mounted on unmanned aerial vehicles (UAVs) can further enhance the MEC performance by decreasing access distance between servers and IoT devices (IDs) \cite{YZeng:CM16, NHMotlagh:IoTJ16, BLi:IoTJ19}. \textcolor{black}{The integration of IoT devices with the MEC enables more responsive and efficient handling of data, addressing latency-sensitive and bandwidth-intensive applications such as smart cities, healthcare monitoring, and industrial automation \cite{ZLiu_IoT:23, HZhou_IoT:23, HXie_IoT:24}.}
On the other hand, the mobility of UAVs incurs time-varying system dynamics, e.g., highly fluctuating propagation statistics. To tackle this difficulty, there have been studies to utilize the deep reinforcement learning (DRL) approaches \cite{YLiu_SADRL:TVT20, LZhang_SADRL:Access21, QLiu_SADRL:TVT20, Wang_SADRL:WCMCC, Wang_SADRL:Access19, Peng_SADRL:WCNC, LWang_SADRL:TMC22, Hwang22:Globecom22, Zchen_SADRL:EURA20, Fsong_SADRL:TMC23, Park:TVT23_MAPPO, BZhang:IoTJ23_early_MAPPO, BLi:ITGCN23_MAPPO, Hwang22:WCom22_MADDPG, LWang:Cog22_MADDPG, JYang:NetSer21_MADDPG, Gao:TVT21_MADDPG, HPeng:JSAC21_MADDPG, Zhao:TWC22_MATD3, CHLiu:JSAC23_MATD3}. 

Centralized DRL frameworks were developed for optimizing the trajectory of UAV servers and resource allocation strategies \cite{YLiu_SADRL:TVT20, Wang_SADRL:Access19, Wang_SADRL:WCMCC, QLiu_SADRL:TVT20, Peng_SADRL:WCNC, LZhang_SADRL:Access21, LWang_SADRL:TMC22, Hwang22:Globecom22, Zchen_SADRL:EURA20, Fsong_SADRL:TMC23}. Deep Q-network (DQN) methods in \cite{YLiu_SADRL:TVT20, LZhang_SADRL:Access21, Wang_SADRL:WCMCC, Wang_SADRL:Access19, QLiu_SADRL:TVT20, Peng_SADRL:WCNC} characterized these optimization variables as actions at an agent, i.e., UAV servers and IDs, which is realized by neural networks (NNs). Since the DQN is confined to discrete action spaces, it fails to identify continuous-valued UAV trajectory. Thus, suitable DRL schemes which can handle continuous action space were introduced such as deep deterministic policy gradient (DDPG), twin delayed DDPG (TD3), and proximal policy optimization (PPO) \cite{LWang_SADRL:TMC22, Hwang22:Globecom22, Zchen_SADRL:EURA20, Fsong_SADRL:TMC23} based on an actor-critic architecture. The actor NN generates actions of UAV agents and ID agents, e.g., trajectory and resource allocation solutions, whereas the critic NN evaluates the effectiveness of the actor NN. By doing so, continuous-valued optimization variables can be successfully obtained for the UAV-based MEC systems.

\subsection{Motivations}

In practical UAV-aided MEC networks, the distributed operation is desirable so that observing states of environments and inferring actions for individual UAVs and IDs can be split across multiple UAVs and IDs. Existing centralized DRL methods \cite{LWang_SADRL:TMC22, Hwang22:Globecom22, Zchen_SADRL:EURA20, Fsong_SADRL:TMC23} combine all UAVs and IDs as a single agent to handle states and actions. Such a single-agent deep reinforcement learning (SADRL) approach brings state collection and decision-making processes in the central manner, which are infeasible for supporting a massive number of IDs. Also, a sole actor-critic NN can only be trained to a certain MEC network with a fixed number of UAVs and IDs, and thus it cannot be straightforwardly applied to larger MEC systems.

One viable approach for distributed UAV-aided MEC systems is multi-agent DRL (MADRL) \cite{Park:TVT23_MAPPO, BZhang:IoTJ23_early_MAPPO, BLi:ITGCN23_MAPPO, Hwang22:WCom22_MADDPG, LWang:Cog22_MADDPG, JYang:NetSer21_MADDPG, Gao:TVT21_MADDPG, HPeng:JSAC21_MADDPG, Zhao:TWC22_MATD3, CHLiu:JSAC23_MATD3}, which adopts multi-agent partially observable Markov decision processes (POMDP). \textcolor{black}{Computing nodes at UAV servers and ground IDs are interpreted as agents which identify their actions based on partial observations of the overall MEC systems, e.g., channel state information, task volume, and locations. Various MADRL techniques such as the MAPPO \cite{Park:TVT23_MAPPO, BZhang:IoTJ23_early_MAPPO, BLi:ITGCN23_MAPPO}, MADDPG \cite{Hwang22:WCom22_MADDPG, LWang:Cog22_MADDPG, JYang:NetSer21_MADDPG, Gao:TVT21_MADDPG, HPeng:JSAC21_MADDPG} and MATD3 \cite{Zhao:TWC22_MATD3, CHLiu:JSAC23_MATD3} have shown their effectiveness for the decentralized management of UAVs and IDs.}

\textcolor{black}{As the multi-agent POMDP formalism lacks full knowledge of the entire MEC system, a coordination strategy among agents should be employed so that they can estimate the overall state from partial observations. Two popular methods include implicit coordination through reward shaping \cite{Park:TVT23_MAPPO, BZhang:IoTJ23_early_MAPPO, BLi:ITGCN23_MAPPO, Hwang22:WCom22_MADDPG, LWang:Cog22_MADDPG, JYang:NetSer21_MADDPG, Gao:TVT21_MADDPG, HPeng:JSAC21_MADDPG} and exploit coordination through observation shaping \cite{Park:TVT23_MAPPO, BZhang:IoTJ23_early_MAPPO, BLi:ITGCN23_MAPPO, Hwang22:WCom22_MADDPG, LWang:Cog22_MADDPG, JYang:NetSer21_MADDPG, Gao:TVT21_MADDPG, HPeng:JSAC21_MADDPG, Zhao:TWC22_MATD3, CHLiu:JSAC23_MATD3}. The former trains all the agents jointly using carefully designed reward functions such that individual agents can implicitly infer the knowledge of other agents in the training. Since agents can cooperate in the training step via long-term statistics, trained actor NNs are not able to accommodate highly fluctuating environments. In contrast, the latter allows agents to share messages, which are normally given as subsets or manipulations of partial observations. These messages can be conveyed through reliable control channels. Such an explicit message exchange mechanism can adapt to immediate environment changes at the expense of increased communication overhead.}

Both approaches have proven to succeed in various configurations of UAV-based MEC networks. However, these approaches generally require time-consuming trial-and-error validation processes to check the feasibility of all possible combinations of partial observations and rewards. This challenge may become prohibitive as the dimension of observation statistics increases. Moreover, man-made agent coordination policies are not guaranteed to achieve good performance since input features of actor NNs normally resort to manual optimization. For these reasons, it is necessary to develop a new cooperative MADRL framework that can autonomously determine interaction strategies among agents, in particular, communication messages by leveraging NNs.

\subsection{Contributions and Organization}
\textcolor{black}{This paper investigates a cooperative MADRL for UAV-aided MEC networks where NNs at multiple agents determine their decision-making and coordination policies autonomously. We aim at minimizing the total energy consumption of ground IDs by offloading their computational tasks to a mobile edge server mounted on a UAV. This poses joint optimization of the trajectory and computing resource allocation of the UAV along with offloading decisions at IDs. These optimization variables need to be computed by the UAV and IDs individually. The considered problem is classified as a multi-agent POMDP where UAV and ID agents collaboratively identify their action variables only with partial observations. Compared to existing MADRL frameworks which require a handcraft design of UAV-ID coordination, the proposed scheme can handle highly fluctuating and heterogeneous network dynamics.}

\textcolor{black}{In this paper, we propose a novel cooperative multi-agent DDPG (C-MADDPG) framework where NNs at UAV and ID agents determine their policies by cooperating with other agents. Our system is first formulated as a cooperative multi-agent POMDP task which constructs two different actions for individual agents, namely, solution actions and message actions. The solution actions include optimization variables at agents, e.g., trajectory and resource allocation variables at the UAV agent and offloading variables at the ID agents. In addition, the message actions indicate communication messages to be exchanged among the UAV and ID agents. For effective information exchange, the proposed design establishes interactions in uplink (ID-to-UAV) and downlink (UAV-to-ID). Thus, along with the solution actions, the uplink message actions at IDs and the downlink message actions at the UAV are regarded as actions taken by actor NNs. This is a distinct feature of the proposed framework compared to conventional MADDPG approaches \cite{RLowe_MADDPG:Neur2017} where communication messages should be designed synthetically in advance and are fixed in all episodes of training steps.}

In the proposed C-MADDPG, we build a solution actor NN and a message actor NN. In the inference stage, the UAV and ID agents calculate their solution actions by using the solution actor NN. Cooperative inference among these actor NNs establishes uplink-downlink coordination among the UAV and ID agents. To send encoded statistics of partial observations to the UAV agent, the ID agents utilize their message actor NNs that generate uplink message actions. These uplink message actions become an input to the message actor NN at the UAV agent, which creates the downlink message actions intended for individual ID agents. Once the coordination is completed, the UAV and ID agents calculate their solution actions by using the solution actor NNs. In this decision-making process, the message actions are leveraged as side inputs to the solution actor NNs so that the UAV and ID agents can collaboratively decide their optimization variables. Since the proposed cooperative inference does not need any centralized operations, we can construct decentralized optimization solutions for practical UAV-aided MEC networks.

\textcolor{black}{In the proposed cooperative actor NN architecture, the UAV agent aggregates the uplink message actions sent by all ID agents. Therefore, the input dimension of actor NNs at the UAV agent, in particular, the message actor NN, scales with the number of IDs. For this reason, a naive NN structure whose input and output dimensions are fixed leads to poor generalization ability with respect to the ID population. To achieve the scalability, we exploit the concept of the graph attention network (GAT) \cite{GAT18:GAT18}, and adopt the parameter sharing technique where all ID agents utilize the identical actor NN. By doing so, the entire inference becomes independent of the number of IDs and can achieve the scalability.}

The training process of the proposed C-MADDPG requests joint optimization of solution actor NNs, message actor NNs, and critic NNs. To this end, we adopt the centralized training decentralized execution (CTDE) strategy where all actor NNs are trained in an end-to-end manner under the supervision of the critic NN. The trained actor NNs are then deployed to the UAV and IDs for real-time decentralized inference. Consequently, the proposed parameter sharing policy can be implemented without additional communication overheads in the inference step. To further improve the scalablity of actor NNs, we develop a joint training process which leverages several episodes with the arbitrary number of ID agents. Also, we employ a random masking strategy that stochastically prunes input features of the critic NN. Numerical results validate the generalization ability of the proposed scheme and demonstrate the effectiveness of the proposed C-MADDPG framework over existing approaches.

\textcolor{black}{The contributions of this
paper are summarized as follows:}
\begin{itemize}
    \item \textcolor{black}{We propose a novel C-MADDPG framework which establishes self-organizing coordination strategies among UAV and ID agents. Compared to existing MADRL methods \cite{Park:TVT23_MAPPO, BZhang:IoTJ23_early_MAPPO, BLi:ITGCN23_MAPPO, Hwang22:WCom22_MADDPG, LWang:Cog22_MADDPG, JYang:NetSer21_MADDPG, Gao:TVT21_MADDPG, HPeng:JSAC21_MADDPG, Zhao:TWC22_MATD3, CHLiu:JSAC23_MATD3} which design agent coordination messages manually, the proposed approach exploits message actor NNs to allow autonomous RL operations. This framework generates task-oriented agent interaction protocols that are optimized to enhance the expected reward function. Consequently, the proposed method does not require a handcrafted design of observations and rewards. 
    }
    \item \textcolor{black}{In practical UAV-aided MEC networks, the number of IDs may vary from time to time. This requests actor NNs whose inference calculations can be performed independent of the ID populations. To this end, we develop scalable actor NNs based on the parameter sharing strategy where all IDs leverage the identical NNs. However, a straightforward extension of the parameter sharing entails indistinguishable messages at all IDs. To address this difficulty, the GAT mechanism is employed which evaluates the importance of individual IDs and the resulting actor NNs successfully achieve the scalability to the number of IDs.
    }
    \item \textcolor{black}{To further enhance the scalability, the training mechanism of the proposed C-MADDPG should be carefully designed such that shared actor NNs can observe various MEC configurations with arbitrary ID populations. Such randomized samples can be generated by using masking operations. We randomly prune IDs in the training step so that the actor NNs can be optimized over different MEC networks. This strategy helps the NNs to learn an efficient decentralized decision-making policy for arbitrary given ID populations.}
\end{itemize}

\textcolor{black}{The rest of the paper is organized as follows: Section \ref{sec:rel} offers an overview of recent works on MADRL-based UAV MEC systems. 
In Section \uppercase\expandafter{\romannumeral3}, we describe a system model and formulate an optimization problem. Section \uppercase\expandafter{\romannumeral4} provides a cooperative POMDP formulation. The proposed actor structure and its cooperative inference are presented in Section \uppercase\expandafter{\romannumeral5}. In Section \uppercase\expandafter{\romannumeral6}, a joint training policy is introduced, and the performance evaluations are shown in Section \uppercase\expandafter{\romannumeral7}. Finally, the paper is terminated with concluding remarks in Section \uppercase\expandafter{\romannumeral8}.}

\section{Related Works}\label{sec:rel}

\textcolor{black}{
Centralized single-agent DRL (SADRL) has been developed to tackle various optimization problems in UAV-aided MEC networks \cite{YLiu_SADRL:TVT20, Wang_SADRL:Access19, Wang_SADRL:WCMCC, QLiu_SADRL:TVT20, Peng_SADRL:WCNC, LZhang_SADRL:Access21, LWang_SADRL:TMC22, Hwang22:Globecom22, Zchen_SADRL:EURA20, Fsong_SADRL:TMC23}. In \cite{YLiu_SADRL:TVT20}, the utility maximization problem of a UAV server was studied based on the DQN framework. The energy consumption was minimized by scheduling the offloading and the position of the UAV using DQN \cite{LZhang_SADRL:Access21}. To obtain continuous-valued UAV trajectories, the DDPG and PPO were taken into account \cite{LWang_SADRL:TMC22, Hwang22:Globecom22, Zchen_SADRL:EURA20, Fsong_SADRL:TMC23}. The DDPG method \cite{LWang_SADRL:TMC22} determined the UAV trajectory to minimize the energy consumption of IDs. The minimization of task completion latency and energy consumption was considered in \cite{Fsong_SADRL:TMC23}.}

\textcolor{black}{
To reduce the complexity of the SADRL, there has been a recent paradigm shift towards the MADRL \cite{Park:TVT23_MAPPO, BZhang:IoTJ23_early_MAPPO, BLi:ITGCN23_MAPPO, Hwang22:WCom22_MADDPG, LWang:Cog22_MADDPG,  Gao:TVT21_MADDPG, JYang:NetSer21_MADDPG, HPeng:JSAC21_MADDPG, Zhao:TWC22_MATD3, CHLiu:JSAC23_MATD3}. These methods offer an effective mechanism for dealing with cooperative or competitive interactions within intricate environments. 
In \cite{Park:TVT23_MAPPO}, the deployment of MEC-powered UAVs was optimized for sub-THz communication. A modified MAPPO was proposed in \cite{BZhang:IoTJ23_early_MAPPO} to handle the energy consumption minimization problem effectively. A joint optimization problem of precoder, trajectory, and ID association was solved in an integrated sensing and communication network \cite{BLi:ITGCN23_MAPPO}. A fairness maximization problem was examined in \cite{LWang:Cog22_MADDPG} by optimizing trajectory and offloading decisions. In \cite{JYang:NetSer21_MADDPG}, total delay and energy consumption were minimized by adopting the stochastic game. The ratio of the transmission rate to the energy consumption of a UAV was optimized in \cite{Gao:TVT21_MADDPG} by combining the game theory and MADDPG. The authors in \cite{HPeng:JSAC21_MADDPG} maximized the number of offloaded tasks while meeting heterogenous quality-of-service requirements. The multi-UAV multi-clouds task offloading problems were addressed in \cite{Zhao:TWC22_MATD3}.}

\textcolor{black}{These MADRL approaches \cite{Park:TVT23_MAPPO, BZhang:IoTJ23_early_MAPPO, BLi:ITGCN23_MAPPO, Hwang22:WCom22_MADDPG, LWang:Cog22_MADDPG, JYang:NetSer21_MADDPG, Gao:TVT21_MADDPG, HPeng:JSAC21_MADDPG, Zhao:TWC22_MATD3, CHLiu:JSAC23_MATD3} generally require exhaustive search processes for identifying efficient rewards and observations heuristically.}
\textcolor{black}{
Weighted sum reward functions were considered in \cite{BZhang:IoTJ23_early_MAPPO, BLi:ITGCN23_MAPPO, Gao:TVT21_MADDPG} where the optimized weights should be found numerically. The work in \cite{JYang:NetSer21_MADDPG} designed the reward as the difference between local and edge computing cost, whereas \cite{HPeng:JSAC21_MADDPG} employed the smoothen objective function as the reward. Also, in \cite{Zhao:TWC22_MATD3} and \cite{CHLiu:JSAC23_MATD3}, the UAV agents exchange their current locations, and these communication messages contribute to partial observation inputs for actor NNs. These man-made agent coordination polices are computationally inefficient and even become infeasible for practical UAV MEC systems with a number of heterogeneous observations and actions.} 

\section{Network Model} \label{sec:system and problem}

\begin{figure}[t]
\begin{center}
\includegraphics[width=\FigureWidth]{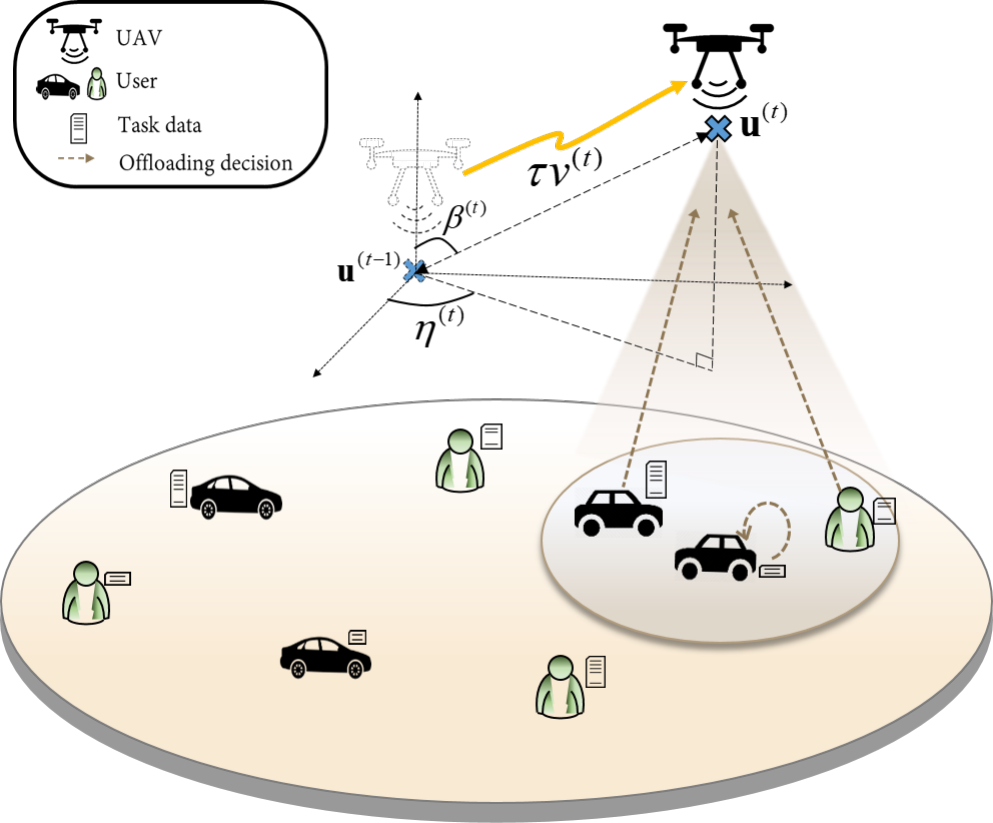}
\end{center}
\caption{UAV-assisted MEC system model}
\label{figure:system}
\end{figure}

As illustrated in Fig. \ref{figure:system}, we consider a UAV-assisted MEC system where the UAV server flies over the network area to offer computation offloading services for $N$ mobile IDs at the ground. A time-slotted MEC protocol is adopted where the system block is divided into $T$ time slots. Let $\mathcal{N}\triangleq\{1,\cdots,N\}$ be the index set of IDs. ID $j$ ($\forall j\in\mathcal{N}$) desires to handle its computational task of size $I_{j}$ bits within one system block consisting of $T$ time slots. 

\subsection{Mobility Model}
Let \textcolor{black}{$\mathbf{q}_{j}^{(t)}=(q_{x,j}^{(t)},q_{y,j}^{(t)},0)\in\mathbb{R}^{3}$} be the 3D Cartesian coordinate of ID $j$ at time slot $t$ ($t=1,\cdots,T$). Mobile IDs change their positions time to time according to predefined missions. The randomness in ID positions can be modeled by the Gauss-Markov process \cite{SBatabyal:15Surveytuts}. 
\textcolor{black}{At time slot $t$, the location of ID $j$ is written by
\begin{subequations}
\begin{align}
    q_{x,j}^{(t)} &= q_{x,j}^{(t-1)} + \tau v_{j}^{(t)}\cos{o_{j}^{(t)}},\\
    q_{y,j}^{(t)} &= q_{y,j}^{(t-1)} + \tau v_{j}^{(t)}\sin{o_{j}^{(t)}},
\end{align}
\end{subequations}where $\tau$ represents the duration of a time slot and the speed $v_{j}^{(t)}$ and $o_{j}^{(t)}$ indicate speed and moving direction, respectively. Here, $v_{j}^{(t)}$ and $o_{j}^{(t)}$ are updated as
\begin{subequations}
\begin{eqnarray}\label{eq:MUs-mobility}
v_{j}^{(t)} = \kappa_{v}v_{j}^{(t-1)} + (1-\kappa_{v})\bar{v}_{j} + \sqrt{1-\kappa_{v}^2} \Phi_{v}, \\
o_{j}^{(t)} = \kappa_{o}o_{j}^{(t-1)} + (1-\kappa_{o})\bar{o}_{j} + \sqrt{\smash[b]{1-\kappa_{o}^2}} \Phi_{o},
\end{eqnarray}
\end{subequations}where $\kappa_{v}\in[0,1]$ and $\kappa_{o}\in[0,1]$ stand for the memory factors and $\bar{v}_{j}$ and $\bar{o}_{j}$ are the average speed and direction of ID $j$, respectively. The independent Gaussian random variables $\Phi_{v} \sim N(0,\varsigma _{v}^2)$ and $\Phi_{o} \sim N(0,\varsigma_{o}^2)$ characterize the randomness of the ID mobility.} 
In the meantime, the UAV trajectory is optimized to enhance the MEC performance. Let us define $\beta^{(t)}\in[0,2\pi]$ and $\eta^{(t)}\in[0,\pi]$ respectively as the azimuth angle and the elevation angle of the UAV at time slot $t$. Then, the moving direction of the UAV $\mathbf{\Delta}^{(t)}\in\mathbb{R}^{3}$ can be expressed as
\textcolor{black}{\begin{align}
\mathbf{\Delta}^{(t)}=(\sin{\beta^{(t)}}\cos{\eta^{(t)}}, \sin{\beta^{(t)}}\sin{\eta^{(t)}}, \cos{\beta^{(t)}}).
\end{align}}
As a result, the 3D location vector of the UAV is obtained as
\begin{align}\label{eq:location_UAV}
\mathbf{u}^{(t)} = \mathbf{u}^{(t-1)} + \tau v^{(t)}\mathbf{\Delta}^{(t)},
\end{align}
where $v^{(t)}\in[0,v_{\max}]$ equals the UAV velocity with $v_{\max}$ being the maximum speed constraint.

\subsection{Channel Model}
We define $P_{j}^{(t)}$ as the line of sight (LoS) probability given by \cite{AAl:WCL14} 
\begin{align}
    P_{j}^{(t)} = \frac{1}{{1 + K_{1}\exp{(-K_{2}[\nu_{j}^{(t)}-K_{1}])}}},\label{eq:Pmnt}
\end{align}
where $K_{1}$ and $K_{2}$ are constants on the propagation environment and $\nu_{j}^{(t)}$ equals the elevation angle between the UAV and ID $j$.
According to the air-to-ground propagation model \cite{MMozaffari:TWC16, YZeng:TWC19}, the large-scale channel gain between the UAV and ID $j$ can be written by
\begin{align}
    h_{j}^{(t)}=\frac{\|\mathbf{u}^{(t)}-\mathbf{q}_{j}^{(t)}\|^{-\alpha}}{\rho_{0}(P_{j}^{(t)}\chi_{\text{LoS}}+(1-P_{j}^{(t)})\chi_{\text{NLoS}})},
\end{align}
where $\mathbf{q}_{j}^{(t)}$ is the 3D location vector of ID $j$, $\alpha$ represents the path loss exponent, $\rho_{0}$ indicates the reference path loss, and $\chi_{\text{LoS}}$ and $\chi_{\text{NLoS}}$ ($\chi_{\text{NLoS}}>\chi_{\text{LoS}}>1$) respectively account for the path loss of the LoS and non-LoS cases. 

We employ the time division duplexing protocol where uplink and downlink communication are realized over the reciprocal channel. 
The uplink and downlink rates are respectively expressed by
\begin{align}\label{eq:rate}
    R_{u,j}^{(t)}&=\! \frac{B}{N}{\log _2}\bigg( {1 \!+\! \frac{{Np_{u}h_{j}^{(t)}}}{{BN_{0}}}} \bigg),\\
 R_{d,j}^{(t)}& =\! \frac{B}{N}{\log _2}\bigg( {1 \!+\! \frac{{Np_{d}h_{j}^{(t)}}}{{BN_{0}}}} \bigg),
\end{align}
where $p_{u}$ and $p_{d}$ equal the uplink and downlink transmit power at the IDs and the UAV, respectively, $B$ denotes the total bandwidth and $N_{0}$ stands for the noise power.

\subsection{Offloading Process}
ID $j$ splits its task into $T$ subtasks each having $I_{j}/T$ bits. Each subtask is subject to be completed within one time slot of duration $\tau$. At the beginning of each time slot, the IDs determine their task offloading policies based on the partial offloading protocol. ID $j$ offloads $\lambda_{j}^{(t)}\in[0,1]$ portion of $I_{j}/T$ bits to the UAV server, whereas the remaining part $1-\lambda_{j}^{(t)}$ is processed locally. The energy consumption $E_{l_{j}}^{(t)}$ required for the local processing of $(1-\lambda_{j}^{(t)})I_{j}/T$ bits is written as~\cite{YWang:TCOM16}
\begin{align}\label{eq:local_energy}
E_{l_{j}}^{(t)} = \vartheta \frac{(C(1-\lambda_{j}^{(t)})I_{j})^3}{\tau^{2}T^{3}},
\end{align}
where the constants $\vartheta$ and $C$ account for the hardware efficiency and the computational complexity, respectively.

To offload $\lambda_{j}^{(t)}\frac{{I}_{j}}{T}$ bits to the UAV, the communication energy $E_{o_{j}}^{(t)}$ of ID $j$ is given by
\begin{align}\label{eq:offload_energy}
E_{o_{j}}^{(t)} = \frac{p_{u}\lambda_{j}^{(t)}I_{j}}{R_{u,j}^{(t)}T}.
\end{align}
The computation capacity of the UAV is limited by the maximum CPU frequency $f_{\max}$. For parallel computations, a virtual machine (VM) \cite{Goldberg:VM74} with the CPU frequency $f_{j}^{(t)}$ is dedicated to processing the task offloaded from ID $j$. This incurs the sum CPU frequency constraint as
\begin{align}
\sum\limits_{j=1}^{N} {f_{j}^{(t)} \le {f_{\max}}}, \forall t. \label{const:freque_const}
\end{align}

\textcolor{black}{The latency of the offloading procedure at ID $j$ comprises delays in the uplink task offloading from ID $j$ to the UAV, task computation at the UAV, and downlink transmission from the UAV to ID $j$ as
\begin{align}
L_{o_{j}}^{(t)}=\frac{\lambda_{j}^{(t)}I_{j}}{R_{u,j}^{(t)}T} + \frac{C{\lambda_{j}^{(t)}I_{j}}}{{f_{j}^{(t)}T}} + \frac{\delta \lambda_{j}^{(t)}I_{j}}{R_{d,j}^{(t)}T}, \label{eq:offload_latency}
\end{align} 
where the first term represents the uplink transmission delay of the offloaded task of size $\lambda_{j}^{(t)}I_{j}$ with $R_{u,j}^{(t)}$ bits/sec, and the second term indicates the computation latency with the CPU frequency $f_{j}^{(t)}$ cycles/sec, and the third term quantifies the downlink transmission delay for broadcasting the task of size $\delta \lambda_{j}^{(t)}I_{j}$ with $R_{d,j}^{(t)}$ bits/sec with the constant $\delta$ being the ratio of output to input task sizes.}
Finally, the latency constraint is imposed as
\begin{align}
L_{o_{j}}^{(t)} \le {\tau}, \forall j, t. \label{const:latency_const}
\end{align}

\subsection{Problem Description}
We aim at minimizing the total energy consumption of all IDs through the joint optimization of the UAV trajectory $\mathbf{U}=\{v^{(t)},\eta^{(t)},\beta^{(t)},\forall t \}$, computing resource allocation $\mathbf{F}=\{f_{j}^{(t)},\forall j,t\}$, and offloading ratio $\mathbf{\Lambda}=\{\lambda_{j}^{(t)},\forall j,t\}$. The total energy minimization problem can be formulated as
\begin{subequations}\label{eq:Problem}
\begin{align}
&\min_{\substack{\mathbf{U},\mathbf{F},\mathbf{\Lambda}}} \quad \frac{1}{T}\sum\limits_{t=1}^{T}\sum\limits_{j=1}^{N}E_{l_{j}}^{(t)} + E_{o_{j}}^{(t)} \nonumber \\
\text{s.t.} & \quad v^{(t)}\in[0,v_{\max}],\ \eta^{(t)}\in[0,2\pi),\ \beta^{(t)}\in[0,\pi], \label{eq:const-q}\\
\quad & \quad f_{j}^{(t)} \geq 0, \label{eq:const-f} \\
\quad & \quad \lambda_{j}^{(t)} \in[0,1],\label{eq:const-beta} \\
\quad & \quad (\ref{const:freque_const})\ \text{and}\ (\ref{const:latency_const}). 
\label{const9}
\end{align}
\end{subequations}
The above problem is a nonconvex problem due to the latency constraint and the objective function. \textcolor{black}{Existing MADRL methods \cite{Park:TVT23_MAPPO, BZhang:IoTJ23_early_MAPPO, BLi:ITGCN23_MAPPO, Hwang22:WCom22_MADDPG, LWang:Cog22_MADDPG,  Gao:TVT21_MADDPG, JYang:NetSer21_MADDPG, HPeng:JSAC21_MADDPG, Zhao:TWC22_MATD3, CHLiu:JSAC23_MATD3} resort to computationally demanding exhaustive search processes for designing agent interaction mechanisms. To overcome this difficulty, we propose a novel cooperative MADRL framework that identifies optimization variables and coordination strategies autonomously by using NNs.}

\section{Cooperative Multi-Agent POMDP Formulation\label{sec:MDP formulation}}

We introduce a C-MADDPG scheme for addressing $(\mathbf{P})$ where the UAV and IDs are realized as individual agents taking their decision variables. Separate UAV and IDs independently interact with the environment, i.e., the MEC network. This leads to a cooperative POMDP formulation where each agent can only access to partial knowledge on the current environment. In what follows, we transform $(\mathbf{P})$ into the multi-agent POMDP task consisting of states, observations, actions, and rewards.

1) Observations and States

We denote the UAV as agent 0 and ID $j$ as agent $j$. Let $\tilde{\mathcal{N}}\triangleq\{0\}\bigcup\mathcal{N}$ be the set of all agents. The partial observation $o_{j}^{(t)}$ of agent $j\in\tilde{\mathcal{N}}$ at time slot $t$ consists of information about the entire MEC network that can be observed by agent $j$. For the UAV agent, the observation $o_{0}^{(t)}$ is set to its previous location $\mathbf{u}^{(t-1)}$ as
\begin{align}
    o_{0}^{(t)}=\mathbf{u}^{(t-1)}. \label{eq:o0}
\end{align}
In contrast, ID agent $j$ forms its observation $o_{j}^{(t)}$ as 
\begin{align}
    o_{j}^{(t)}=\left\{\mathbf{q}_{j}^{(t-1)}, (1-\lambda_{j}^{(t-1)})\frac{I_{j}}{T}, \lambda_{j}^{(t-1)}\frac{I_{j}}{T}, I_{j}, R_{u,j}^{(t-1)}\right\}. \label{eq:observation}
\end{align}
As a result, the state of the MEC network $s^{(t)}$ collects all observations as
\begin{align}
    s^{(t)}\triangleq\{o_{j}^{(t)}:\forall j\in\tilde{\mathcal{N}}\}. \label{eq:state}
\end{align}

2) Solution Actions

The solution action $x_{j}^{(t)}$ contains a set of optimization variables identified by agent $j$. As discussed, the solution action $x_{0}^{(t)}$ of the UAV agent receives the trajectory variables as 
\begin{align}
    x_{0}^{(t)}=\{\mathbf{v}^{(t)}, \mathbf{f}^{(t)}\}, \label{eq:sol_act0}
\end{align}
where $\mathbf{v}^{(t)}=\{v^{(t)}, \eta^{(t)}, \beta^{(t)}\}$ is the trajectory variable and $\mathbf{f}^{(t)}\triangleq\{f_{j}^{(t)}:\forall j\in\mathcal{N}\}$ stands for a collection of the CPU frequencies of all IDs. Also, ID agent $j$ obtains its own offloading decision variable $\lambda_{j}^{(t)}$. Thus, the solution $x^{(t)}_{j}$ of ID agent $j$ becomes
\begin{align}
    x^{(t)}_{j}=\lambda_{j}^{(t)}.\label{eq:xjt}
\end{align}

3) Reward

The reward function $r^{(t)}$ evaluates the performance of the MEC network at time slot $t$. Since our aim is to minimize the energy consumption of all IDs, the reward $r^{(t)}$ is set to
\begin{align}
    r^{(t)} = - \sum\limits_{j=1}^{N}(E_{l_{j}}^{(t)}+E_{o_{j}}^{(t)}).\label{eq:reward}
\end{align}

4) Message Actions

The considered POMDP formulation can be addressed by the conventional MADDPG framework \cite{RLowe_MADDPG:Neur2017}. In this approach, agent $j$ ($\forall j\in\tilde{\mathcal{N}}$) is equipped with its own actor NN, which produces the solution action $x_{j}^{(t)}$ from the partial observation $o_{j}^{(t)}$. The major drawback of the conventional MADDPG comes from the limited agent interaction. Since the solution actions of all agents are highly coupled in the UAV-aided MEC networks, coordination among the UAV and IDs is essential for identifying the optimal solution to \textbf{(P)}. Nevertheless, the actor DNN only accepts the partial observation $o_{j}^{(t)}$ as an input, and thus the resulting solution action $x_{j}^{(t)}$ is determined without knowing the observations of other agents.

To cope with this issue, along with the decision processes of the solution actions, we develop a coordination policy among the UAV and ID agents, which can be realized by additional message actions $m_{j}^{(t)}$ ($\forall j\in\tilde{\mathcal{N}}$). The message actions should be designed to encapsulate sufficient statistics of agent $j$ needed for individual decision-making processes at others. Messages of ID agents are shared with the UAV through uplink control channels. Similarly, the UAV multicasts its message action $m_{0}^{(t)}$ to all IDs via downlink control channels. As will be explained, the message actions are determined using additional actor NNs, and the resulting message actors are leveraged as side information to decide the solution actions. Thus, the overall action of agent $j$ consists of both the solution action and message action as
\begin{align}
    a_{j}^{(t)}=\{x_{j}^{(t)},m_{j}^{(t)}\}.
\end{align}

\section{Cooperative Actor Design}\label{sec:Proposed}
\begin{figure*}[t]
\begin{center}
\includegraphics[width=\linewidth, height=3.1in]{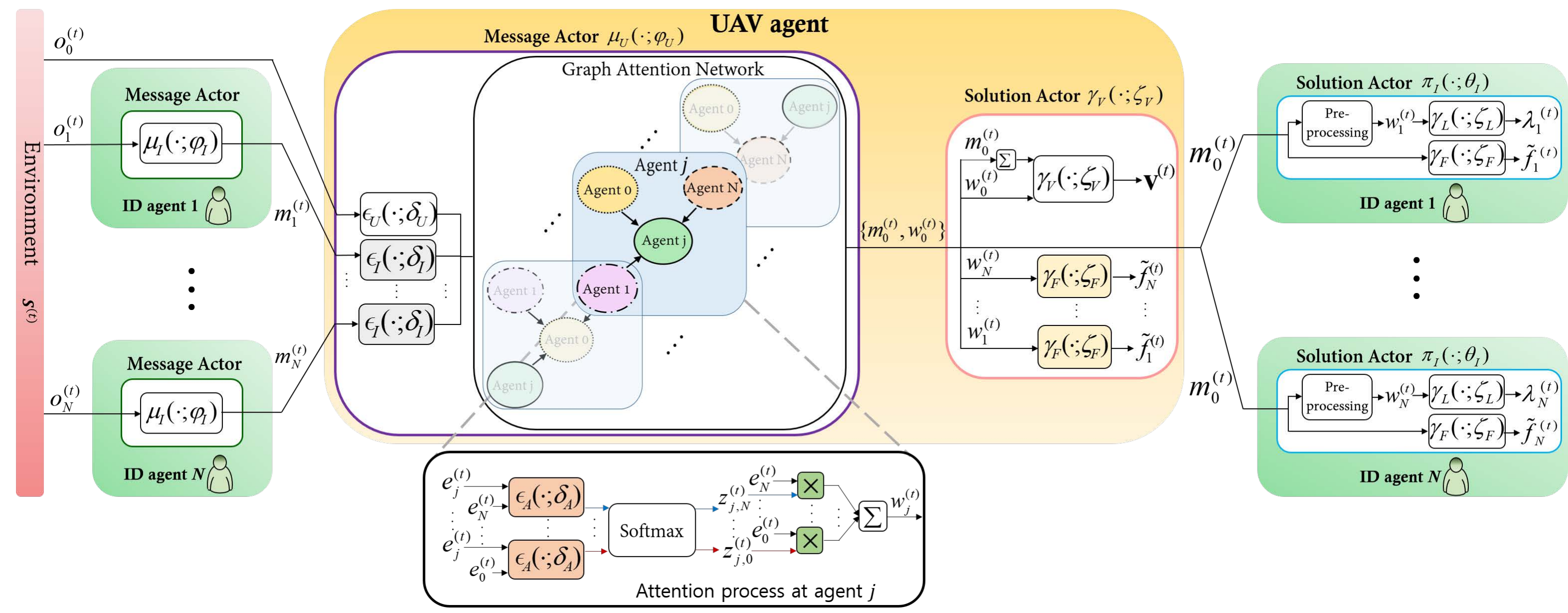}
\end{center}
\caption{Proposed cooperative actor architecture}
\label{figure:actor}
\end{figure*}

The cooperative multi-agent POMDP formulation presented in Section \ref{sec:MDP formulation} readily establishes the C-MADDPG framework to take the solution actions $x^{(t)}_{j}$ ($\forall j\in\tilde{\mathcal{N}}$) using decentralized coordination among the UAV and ID agents. In the C-MADDPG method, this can be achieved by employing actor NNs at individual agents. In order to determine the message actions $m_{j}^{(t)}$ ($\forall j\in\tilde{\mathcal{N}}$), we employ additional actor NNs that produce message action variables. As illustrated in Fig. \ref{figure:actor}, the proposed architecture deploys two types of actors: \textit{message actor} and \textit{solution actor}. The message actors determine agent coordination, whereas the solution actors compute appropriate solution variables based on the received message actions. Such a cooperative actor architecture leads to the joint optimization of two different actions in a goal-oriented manner for maximizing the expected reward value.

The proposed actor design invokes a challenging issue on the scalability with respect to the network size, in particular, the number of IDs $N$. The joint optimization of message actors and solution actors poses the exploding population of trainable parameters that is proportional to $2N$. Such dedicated actor NNs lack the generalization ability to an arbitrary $N$. Also, the sizes of the state $s^{(t)}$ and message actions $\{m^{(t)}_{j}:\forall j\in\mathcal{N}\}$ grow with $N$, which increases the model complexity for critic and actor NNs to handle high-dimensional states and actions. To address these difficulties, in this section, we develop a cooperative and scalable actor structure whose message-generating and solution-optimizing computations become independent of the number of IDs. In what follows, we discuss the inference steps of message actors and solution actors.


\subsection{Message Actors at ID agents}
ID agent $j$ ($\forall j\in\mathcal{N}$) first obtains its message action $m^{(t)}_{j}$ by the message actor NN $\mu_{I}(\cdot;\varphi_{I})$ with trainable parameter $\varphi_{I}$ based on its partial observation $o_{j}^{(t)}$. The message action $m^{(t)}_{j}$ of ID agent $j$ ($\forall j\in\mathcal{N}$) is then expressed as
\begin{align}
    m_{j}^{(t)}=\mu_{I}(o_{j}^{(t)};\varphi_{I}), \label{eq:MU_Mess_gene}
\end{align}
where the identical message actor $\mu_{I}(\cdot;\varphi_{I})$ is employed for all ID agents. Such a parameter sharing policy leads to a scalable structure so that a sole message actor can be universally applied to an arbitrary ID population. 

A set of ID messages $\{m_{j}^{(t)}:\forall j\in\mathcal{N}\}$ are conveyed to the UAV agent through orthogonal uplink control channels. Without loss of the generality, each ID is assumed to be assigned by $M$ frequency resource blocks (RBs) to transmit its message action. To accommodate such a resource constraint, we design $m_{j}^{(t)}$ as an $M$-dimensional vector where the transmission of each element occupies one RB. As a result, total $NM$ RBs are needed for the uplink coordination.

\subsection{Message Actor at UAV agent}

After receiving the messages $\mathbf{m}_{I}^{(t)}\triangleq\{m_{j}^{(t)}:\forall j\in\mathcal{N}\}$, the UAV agent computes its message action $m_{0}^{(t)}$ to be multicast to all IDs through the downlink control channels. This UAV message action encodes the knowledge required for the decision processes of the solution action $x_{j}^{(t)}$ at ID agent $j\in\mathcal{N}$. To this end, the UAV agent combines its observation $o_{0}^{(t)}$ and the group of ID message actions $\mathbf{m}_{I}^{(t)}$. By doing so, the UAV successfully propagates partitioned information of individual IDs to the entire MEC network. Similar to the ID message actions, the UAV message actor $\mu_{U}(\cdot;\varphi_{U})$ with trainable parameter $\varphi_{U}$ is adopted to produce the UAV message action $m_{0}^{(t)}$ as
\begin{align}
m_{0}^{(t)}=\mu_{U}(o_{0}^{(t)},\mathbf{m}_{I}^{(t)};\varphi_{U}). \label{eq:m0}
\end{align}
The dimension of the ID message actions $\mathbf{m}_{I}^{(t)}\in\mathbb{R}^{NM}$ scales with the number of IDs $N$. For this reason, a naive NN architecture, in particular, fully-connected layers, fails to achieve the scalability with respect to $N$.

To overcome this issue, we develop a scalable UAV message actor $\mu_{U}(\cdot;\varphi_{U})$ based on the GAT \cite{GAT18:GAT18}. This framework modifies node interaction policies of the graph neural network (GNN) \cite{GNN:GNN09} such that each node can measure the importance of its neighbors, which is referred to as an attention score. As a result, the generalization capability can be fairly improved without sacrificing the scalability to the node population. 
Thus, the design goal of the message actor \eqref{eq:m0} is to aggregate the observation of the UAV agent $o_{0}^{(t)}$ and $\mathbf{m}_{I}^{(t)}$ based on the importance of individual IDs.

\textcolor{black}{In general, the GAT facilitates multiple layers to extract useful features of input data. To realize such a layered GAT architecture, several communication rounds among the UAV and IDs are necessary to share the results of each GAT iteration. To avoid this issue, we modify conventional multi-iteration GAT architectures to a single-iteration GAT to leverage sole uplink-downlink coordination. Also, since the DRL involves temporal connections of actor NNs in consecutive time slots, our single-iteration GAT design becomes more powerful in the UAV-aided MEC systems.
}

\textcolor{black}{ A key enabler of the proposed GAT approach is to allow the UAV agent to have NN modules for computing the attention scores of all ID agents. By doing so, we can straightforwardly implement the GAT mechanism without propagating latent vectors of hidden layers multiple times.} The UAV agent first extracts a hidden feature of agent $j\in\tilde{\mathcal{N}}$, denoted by $\mathbf{e}_{j}^{(t)}$ of length $E$, as
\begin{align}
    \mathbf{e}_{j}^{(t)}=\begin{cases}
        \epsilon_{U}(o_{0}^{(t)};\delta_{U})\ \text{for} & j=0, \\
        \epsilon_{I}(m_{j}^{(t)};\delta_{I})\ \text{for} & j\in\mathcal{N},
    \end{cases} \label{eq:UAV_fea_ext} 
\end{align}
\textcolor{black}{where $\epsilon_{U}(\cdot;\delta_{U})$ and $\epsilon_{I}(\cdot;\delta_{I})$ indicate the feature extractor NNs of the UAV agent and ID agents, respectively, which are responsible for generating $\mathbf{e}_{j}^{(t)}$ utilized for the computation of the attention scores.} A group of feature vectors $\{\mathbf{e}_{j}^{(t)}:\forall j\in\tilde{\mathcal{N}}\}$ is adopted as an input to the GAT operation. A scalar attention score $z_{j,k}^{(t)}\in[0,1]$ about agent $k$ measured by agent $j$ is calculated as
\begin{align}
    z_{j,k}^{(t)}=\frac{\exp(\epsilon_{A}(\mathbf{e}_{j}^{(t)}, \mathbf{e}_{k}^{(t)};\delta_{A}))}{\sum_{i\in\tilde{\mathcal{N}}}\exp(\epsilon_{A}(\mathbf{e}_{j}^{(t)},\mathbf{e}_{i}^{(t)};\delta_{A}))}, \label{eq:Gattention}
\end{align}
where $\epsilon_{A}(\cdot;\delta_{A})$ stands for the attention NN that evaluates the affinity of two different agents. The attention score $z_{j,k}^{(t)}$ interprets the importance of agent $k$ for the decision-making process at agent $j$. Finally, the output of the GAT for agent $j$ becomes the weighted average of the feature vectors $\{\mathbf{e}_{j}^{(t)}:\forall j\in\tilde{\mathcal{N}}\}$ with coefficients $\{z_{j,k}^{(t)}:\forall j\in\tilde{\mathcal{N}}\}$ as
\begin{align}
    \mathbf{w}_{j}^{(t)}=\sum\limits_{k\in\tilde{\mathcal{N}}}z_{j,k}^{(t)}\mathbf{e}_{k}^{(t)}. \label{eq:DW_Mess_gene}
\end{align}
\textcolor{black}{Notice that in conventional GAT, each agent has its dedicated feature extractor NN. Thus, to obtain the attention score $z_{j,k}^{(t)}$ in \eqref{eq:Gattention}, ID agent $j$ should know all the feature vectors $\mathbf{e}_{k}^{(t)}$ ($\forall k\in\tilde{\mathcal{N}}$) which is not viable with sole uplink cooperation from IDs to the UAV. This can be addressed by allowing the UAV agent to reuse the feature extractor NNs $\epsilon_{I}(\cdot;\delta_{I})$ for all IDs.}

Thanks to the GAT mechanism, $\mathbf{w}_{j}^{(t)}$ encodes sufficient statistics required at agent $j$ to take its solution actions. Thus, they can be utilized as an input of solution actor NNs. As will be discussed, $\mathbf{w}_{0}^{(t)}$ is leveraged internally to find the solution action of the UAV agent $x_{0}^{(t)}$. In contrast, the remaining vectors $\mathbf{w}_{j}^{(t)}$ ($\forall j\in\mathcal{N}$) need to be sent to the associated ID agents. To this end, the UAV message action $m_{0}^{(t)}$ is designed~as
\begin{align}
m_{0}^{(t)} = \{\mathbf{w}_{j}^{(t)}:\forall j\in\mathcal{N}\}. \label{eq:UAV_messa}
\end{align}
The UAV multicasts $m_{0}^{(t)}\in\mathbb{R}^{NE}$ to the IDs, which occupies $NE$ RBs for the downlink coordination.

\subsection{Solution Actor at UAV agent}
The UAV message $m_{0}^{(t)}$ encapsulates the network state $s^{(t)}$, i.e., the set of all observations of the UAV and ID agents. Therefore, it is sufficient for all agents to determine their solution actions by leveraging $m_{0}^{(t)}$ only. 
Let $\pi_{U}(\cdot;\theta_{U})$ be the solution actor NN at the UAV agent with the learnable parameter $\theta_{U}$. The UAV solution action $x_{0}^{(t)}$ is obtained as
\begin{align}
    x_{0}^{(t)}=\pi_{U}(m_{0}^{(t)};\theta_{U}). \label{eq:UAV_actor}
\end{align}
As shown in \eqref{eq:sol_act0}, the solution action of the UAV agent $x_{0}^{(t)}$ contains two types of optimization variables, i.e., the trajectory $\mathbf{v}^{(t)}$ and computing resource allocation $\mathbf{f}^{(t)}$. To yield such heterogeneous actions, the solution actor $\pi_{U}(\cdot;\theta_{U})$ comprises two component NNs $\gamma_{V}(\cdot;\zeta_{V})$ and $\gamma_{F}(\cdot;\zeta_{F})$ for calculating $\mathbf{v}^{(t)}$ and $\mathbf{f}^{(t)}$, respectively. Then, the trainable parameter set of the solution actor NN of the UAV becomes $\theta_{U}=\{\zeta_{V},\zeta_{F}\}$.

The trajectory variable is computed as
\begin{align}
\mathbf{v}^{(t)}=\gamma_{V}\left(\mathbf{w}_{0}^{(t)},\sum_{j\in\mathcal{N}}\mathbf{w}_{j}^{(t)};\zeta_{V}\right), \label{eq:UAV_actor_trj}
\end{align}
where an input is a concatenation of the UAV information vector $\mathbf{w}_{0}^{(t)}\in\mathbb{R}^{E}$ from the GAT \eqref{eq:Gattention} and the sum of the ID information vectors $\sum_{j\in\mathcal{N}}\mathbf{w}_{j}^{(t)}\in\mathbb{R}^{E}$. Since the dimension of these input vectors is independent of $N$, \eqref{eq:UAV_actor_trj} preserves the scalability with respect to the ID population. The distinct input $\mathbf{w}_{0}^{(t)}$ helps the NN $\gamma_{V}(\cdot;\zeta_{V})$ distinguish the UAV information vector with those of the IDs $\sum_{j\in\mathcal{N}}\mathbf{w}_{j}$. Consequently, we can successfully produce the UAV-specific trajectory action $\mathbf{v}^{(t)}$ based on the aggregated ID information $\sum_{j\in\mathcal{N}}\mathbf{w}_{j}$.

Next, to determine the computing resource allocation $\mathbf{f}^{(t)}$, a sole NN produces each CPU frequency variable $f_{j}^{(t)}$ based on the associated ID agent information vector $\mathbf{w}_{j}^{(t)}$. To this end, we first calculate an intermediate value $\tilde{f}_{j}^{(t)}$ as
\begin{align}
    \tilde{f}_{j}^{(t)}=\gamma_{F}(\mathbf{w}_{j}^{(t)};\zeta_{F}), \label{eq:UAV_resuorce_allo}
\end{align}
where the output activation of $\gamma_{F}(\cdot;\zeta_{F})$ is set to the rectified linear unit (ReLU) to yield a nonnegative number $\tilde{f}^{(t)}_{j}$. Upon obtaining all $\tilde{f}_{j}^{(t)}$, the CPU cycle action $f_{j}^{(t)}$ is retrieved as
\begin{align}
    f_{j}^{(t)}=\frac{\tilde{f}_{j}^{(t)}}{\sum_{k\in\mathcal{N}}\tilde{f}_{k}^{(t)}}f_{\max}, \label{eq:f_norm}
\end{align}
which forces $\mathbf{f}^{(t)}$ to be feasible for the computing resource constraint \eqref{const:freque_const}.

\subsection{Solution Actor at ID agent}
From the message action $m_{0}^{(t)}$ from the UAV, ID agent $j$ first recovers its corresponding information vector $\mathbf{w}_{j}^{(t)}$. Let $\pi_{I}(\cdot;\theta_{I})$ be the solution actor NN of the ID agent with parameter $\theta_{I}$. Then, the ID solution action $x_{j}^{(t)}$ in \eqref{eq:xjt}, which equals the offloading ratio $\lambda_{j}^{(t)}$, is expressed as
\begin{align}
    x_{j}^{(t)}=\pi_{I}(\mathbf{w}_{j}^{(t)};\theta_{I}). \label{eq:ID_actor}
\end{align} 
The latency constraint in \eqref{const:latency_const} always becomes feasible if $x_{j}^{(t)}$ lies within a bounded range $[0,\lambda_{\max,j}^{(t)}]$, where an upperbound $\lambda_{\max,j}^{(t)}$ is given by
\begin{align}
    \lambda^{(t)}_{\max,j} = \min\left\{1, \frac{\tau T}{I_{j}}/(\frac{1}{R_{u,j}^{(t)}} + \frac{\delta}{R_{d,j}^{(t)}}+ \frac{C}{f_{j}^{(t)}})\right\}.\label{eq:feasible_lambda}
\end{align}
To guarantee $x_{j}^{(t)}\in[0,\lambda_{\max,j}^{(t)}]$, the output activation of $\pi_{I}(\cdot;\theta_{I})$ is set to the sigmoid function multiplied by $\lambda_{\max,j}^{(t)}$. To calculate $\lambda_{\max,j}^{(t)}$, ID agent $j$ needs to know its CPU frequency allocation $f_{j}^{(t)}$, which is, in fact, determined at the UAV agent. To this end, the IDs are assumed to be equipped with a copy of the UAV actor NN $\gamma_{F}(\cdot;\zeta_{F})$ in \eqref{eq:UAV_resuorce_allo}. This can be achieved by an offline training procedure which optimizes all actor NNs jointly before its real-time inference step, as will be discussed in Section \ref{sec:secV}. Along with the UAV message $m_{0}^{(t)}$ and the normalization operation in \eqref{eq:f_norm}, each ID agent readily gets its CPU frequency $f_{j}^{(t)}$ without incurring additional communication with the UAV and other ID agents.

\begin{algorithm}[ht]
\caption{\textcolor{black}{Cooperative inference of the proposed C-MADDPG}}
\begin{algorithmic}
\NoDo
\State \textit{1. Uplink coordination} 
    \State \quad ID agent $j$ ($\forall j\in \mathcal{N}$) creates $m_{j}^{(t)}$ from \eqref{eq:MU_Mess_gene} and sends it 
    \State \quad to the UAV agent.
\State \textit{2. Downlink coordination}
\State \quad The UAV agent computes $m_{0}^{(t)}$ from \eqref{eq:m0}-\eqref{eq:UAV_messa} and 
\State \quad multicasts it to all ID agents.
\State \textit{3. Decision at UAV agent}
\State\quad The UAV agent calculates the solution action $x_{0}^{(t)}$ from 
\State \quad \eqref{eq:UAV_actor}-\eqref{eq:UAV_resuorce_allo}.
\State \textit{4. Decision at ID agents}
\State \quad Each ID agent $j$ ($\forall j\in \mathcal{N}$) obtains the solution action 
\State \quad $x_{j}^{(t)}$ from \eqref{eq:ID_actor} and \eqref{eq:feasible_lambda}.
\end{algorithmic}
\label{Algorithm 1}
\end{algorithm}

Algorithm \ref{Algorithm 1} summarizes the cooperative inference among the UAV and ID agents of the proposed C-MADDPG framework. 
At the beginning of the time slot, ID agent $j$ individually generates the ID message $m_{j}^{(t)}$ using its message actor NN $\mu_{I}(\cdot;\varphi_{I})$ in $\eqref{eq:MU_Mess_gene}$. The resulting messages are conveyed to the UAV agent through the uplink coordination channel. Next, the UAV agent generates the UAV message $\mathbf{m}_{0}^{(t)}$ based on its message actor NN $\mu_{U}(\cdot;\varphi_{U})$ and multicasts the output message to the ID agents. After this uplink-downlink agent interaction, each agent individually decides its solution action by leveraging the dedicated solution actor NNs $\pi_{U}(\cdot;\theta_{U})$ and $\pi_{I}(\cdot;\theta_{I})$. 

It is important to note that the proposed cooperative inference can be realized in a decentralized manner. Both the message and solution actions can be determined at individual agents based only on their local information, e.g., observation $o_{j}^{(t)}$ and messages $m_{j}^{(t)}$. As a consequence, neither central information collection steps nor central computing units are necessary to implement Algorithm \ref{Algorithm 1}.
\textcolor{black}{So far, we have designed cooperative actor NNs and their decentralized execution mechanisms for establishing the proposed MADDPG framework. In the following sections, we discuss the MADDPG training algorithm for the proposed actor NNs.}

\section{Joint Training Strategy}\label{sec:secV}
We present a joint training policy which optimizes the actor NNs in an end-to-end manner. 
\textcolor{black}{The CTDE strategy is adopted which trains all NNs centrally in an offline manner and dispatches the trained actor NNs to intended nodes for online and decentralized decisions. This approach guarantees that the identical actor NNs are deployed across all IDs in the training step.
} To train the actor NNs, we employ a critic NN $Q(s^{(t)},a^{(t)};\phi)$ with parameter $\phi$ which estimates the Q-value of the state-action pair $(s^{(t)},a^{(t)})$, where $a^{(t)}$ stands for the global action collecting all actions of the UAV and ID agents as
\begin{align}
    a^{(t)}=\{a_{j}^{(t)}:\forall j\in\tilde{\mathcal{N}}\}. \label{eq:action_global}
\end{align}

The training step leverages a relay buffer $\mathcal{M}$ given as
\begin{align}
    \mathcal{M}=\{(s^{(t)},a^{(t)},r^{(t)},s^{(t+1)}):\forall t\}.
    \label{eq:transmition_sample}
\end{align}
To improve the scalability to the ID population $N$, transition samples in \eqref{eq:transmition_sample} are generated over random $N$ uniformly distributed within $[N_{\min}, N_{\max}]$. This enables a versatile computation architecture for the critic NN $Q(\cdot;\phi)$ to handle variable-length state and action inputs, which can be realized by a simple masking operation to the input. More precisely, for each transition sample, we uniformly set the ID population $N\in[N_{\min}, N_{\max}]$ and sample the index set of ID agents randomly, i.e., $\mathcal{N}\subset\mathcal{N}_{\max}\triangleq\{1,\cdots,N_{\max}\}$ with $|\mathcal{N}|=N$. The observations $o_{j}^{(t)}$ for inactive ID agents $j\in\mathcal{N}_{\max}\backslash\mathcal{N}$ are fixed as zero vectors. Then, the state $s^{(t)}$ in \eqref{eq:state} concatenates all observations $o_{j}^{(t)}$ ($\forall j\in\{0\}\bigcup\mathcal{N}_{\max}$) thereby resulting in a masked vector. A similar operation is applied to construct the global action $a^{(t)}$ in \eqref{eq:action_global}. As a consequence, the critic NN $Q(s^{(t)},a^{(t)};\phi)$ designed to process the maximum ID population $N_{\max}$ preserves the scalability to an arbitrary $N$.

\textcolor{black}{At each training iteration, we randomly sample a mini-batch set $\mathcal{B}\subset\mathcal{M}$ from the replay buffer $\mathcal{M}$. Let $b=(s,a,r,s^{\prime})\in\mathcal{B}$ be a particular batch sample where $s^{\prime}$ represents the one-step forward state originated from the current state-action pair $(s,a)$. Also, we define $\psi\triangleq\{\varphi_{U},\theta_{U},\varphi_{I},\theta_{I}\}$ as the parameter set of all actor NNs. 
The actor NNs are optimized to maximize the policy objective function $J(\psi)$ which measures the expected Q-value over the mini-batch samples $b\in\mathcal{B}$ as
\begin{align}
    J(\psi)=\frac{1}{|\mathcal{B}|}\sum_{b\in\mathcal{B}}Q(s, A(s;\psi);\phi), \label{eq:obj_actor}
\end{align}
where $A(\cdot;\psi)$ indicates a group of actor NNs that take the global action $a$ from the current state $s$. }
\textcolor{black}{As a result, the mini-batch stochastic gradient descent (SGD) update strategy of the actor NNs becomes
\begin{align}
    \psi\leftarrow\psi+\eta_{A}\nabla_{\psi}J(\psi), \label{eq:actor_update}
\end{align}
where $\eta_{A}>0$ denotes the learning rate and $\nabla_{z}$ equals the gradient operator with respect to the variable $z$.}

\textcolor{black}{The critic NN $Q(s,a;\phi)$ is obtained to yield a correct Q-value for a given state-action pair $(s,a)$. The corresponding critic loss function $L(\phi)$ is written by
\begin{align}
    L(\phi)\!=
    \!\frac{1}{|\mathcal{B}|}\sum_{b\in\mathcal{B}}
    \!\!\big(y\!-\!Q(s,a;\phi) \big)^{2},\label{eq:loss_critic}
\end{align}
where $y$ stands for the target of the estimated Q-value $Q(s,a;\phi)$ as
\begin{align}
    y= r + \alpha Q(s^{\prime}, A(s^{\prime};\psi^{\prime});\phi^{\prime}),
\end{align}
with $\alpha\in(0,1]$. Then, the critic NN is updated as
\begin{align}
    \phi\leftarrow\phi-\eta_{C}\nabla_{\phi}L(\phi) \label{eq:critic_update}
\end{align}
where $\eta_{C}$ equals the learning rate for the critic NN. In addition, we adopt the soft update strategy for the target actor NN $A(\cdot;\psi^{\prime})$ and target critic NN $Q(\cdot;\phi^{\prime})$ expressed by
\begin{align}
    \psi^{\prime}&\leftarrow\kappa_{A}\psi+(1-\kappa_{A})\psi^{\prime}\label{eq:actor_target_update}\\
    \phi^{\prime}&\leftarrow\kappa_{C}\phi+(1-\kappa_{C})\phi^{\prime}\label{eq:critic_target_update},
\end{align}
where $\kappa_{C}$ and $\kappa_{A}$ are the target update rates.}

\begin{algorithm}[ht]
\caption{\textcolor{black}{Joint training strategy of the proposed C-MADDPG}}
\begin{algorithmic}
\NoDo
\State{Initialize $\psi$, $\phi$, $\psi^{\prime}$, $\phi^{\prime}$ $\sigma^{2}$ and $\rho$.}
\For{episode $e=1,\cdots,E$}
    \State \textcolor{black}{{Initialize the number of IDs $N\in[N_{\min}, N_{\max}]$.}}
    \State \textcolor{black}{{Sample the set of active IDs $\mathcal{N}\subset\mathcal{N}_{\max}$.}}
    \State \textcolor{black}{{Generate an initial state $s^{(1)}$ by masking local}} 
    \State \textcolor{black}{{observations of inactive IDs $\mathcal{N}_{\max}\backslash\mathcal{N}$.}}
    \State Update the exploration noise variance as $\sigma^2\leftarrow\rho\sigma^2$.
    \For{time slot $t=1,\cdots,T$}    
    \State {Calculate $a^{(t)}=A(s^{(t)};\psi)$ from Algorithm \ref{Algorithm 1}.}
    \State Add the exploration noise as $a^{(t)}\leftarrow a^{(t)} +\mathcal{N}(0,\sigma^{2})$.
    \State {\textcolor{black}{Mask local actions of inactive IDs in $a^{(t)}$.}}
    \State {Obtain reward $r^{(t)}$ and the next state $s^{(t+1)}$.}
    \State {Store $(s^{(t)}, a^{(t)}, r^{(t)}, s^{(t+1)})$ in the replay buffer $\mathcal{M}$}
    \State {Sample the mini-batch set $\mathcal{B}\subset\mathcal{M}$.}
    \State {Update the actor NN $\psi$ and the critic NN $\phi$ from}
    \State {\eqref{eq:actor_update} and \eqref{eq:critic_update}.}
    \State {Update the target actor NN $\psi^{\prime}$ and the target critic} 
    \State {NN $\phi^{\prime}$ from \eqref{eq:actor_target_update} and \eqref{eq:critic_target_update}.}
    
    $\textbf{end}$
   \EndFor
   \EndFor
$\textbf{end}$
\end{algorithmic}
\label{Algorithm 2}
\end{algorithm}

Algorithm \ref{Algorithm 2} presents the joint training process of the proposed C-MADDPG framework. 
\textcolor{black}{Unlike the existing MADDPG training algorithm, the proposed training procedure involves scalable learning strategies to enhance the generalization ability of the actor and critic NNs. In the initialization phase, the number of IDs $N$ is generated uniformly within $[N_{\min}, N_{\max}]$. Then, we randomly sample the set of active IDs as $\mathcal{N} \subset \mathcal{N}_{\max} \triangleq \{1, \cdots, N_{\max}\}$ with $|\mathcal{N}| = N$. A masking operation is employed for inactive IDs where the local observations $o_{j}^{(t)}$ for $j \in \mathcal{N}_{\max} \backslash \mathcal{N}$ are replaced with zero vectors. Consequently, the state $s^{(t)}$ contains the observations of the UAV and active IDs. This masking operation is similarly applied to the action $a^{(t)}$ and this can be viewed as an extension of the dropout operation \cite{Dropout:14} developed for the generalization ability. We discard the local observations and local actions of randomly selected inactive IDs. By doing so, ensemble training of the actor and critic NNs is achieved by producing a number of randomized MEC configurations in the training. This is also beneficial for the actor NNs to learn a generic decentralized decision policy with any given $N$.} \textcolor{black}{The hyperparameters are utilized to improve the exploration capability of the actor NNs by adding the Gaussian noise as
\begin{align}
    a^{(t)}=A(s^{(t)};\psi)+n, \label{eq:exploration}
\end{align}
where $n \backsim \mathcal{N}(0,\sigma^{2})$ stands for the Gaussian random variable. The exploration noise helps the actor NN $A(\cdot;\psi)$ choose new actions that have not been experienced in the training, thereby improving the generalization capability. At the beginning of each episode, we set $N$ uniformly over $[N_{\min},N_{\max}]$. An initial state $s$ is randomly generated according to predefined UAV and ID deployment scenarios. Also, we decrease the exploration noise variance $\sigma^{2}$ by the decaying ratio $\rho$. Thus, the power of the exploration noise in \eqref{eq:exploration} is gradually reduced as the training continues.} 

Each episode consists of $T$ time slots. At each time slot, we execute Algorithm 1 to yield the action $a^{(t)}$ for the current state $s^{(t)}$. Then, the exploration noise is injected into the action $a^{(t)}$ as in \eqref{eq:exploration}. The interaction with the MEC network produces the reward $r^{(t)}$ for the state-action pair $(s^{(t)},a^{(t)})$ as well as the next state $s^{(t+1)}$. The resulting transition sample $(s^{(t)}, a^{(t)}, r^{(t)}, s^{(t+1)})$ is stored to the replay buffer $\mathcal{M}$. Next, we randomly sample the mini-batch set $\mathcal{B}$ from the replay buffer $\mathcal{M}$ uniformly, which is utillized for the training of the actor NNs and critic NN based on their update rules in \eqref{eq:actor_update} and \eqref{eq:critic_update}. It is then followed by the modification steps of the target NNs in \eqref{eq:actor_target_update} and \eqref{eq:critic_target_update}. These procedures are repeated for $T$ time slots, which complete one training episode. The entire training step elapses total $E$ episodes.

The proposed joint training algorithm is conducted centrally in an offline manner.  This is due to the critic NN $Q(s,a;\phi)$ which estimates the Q-value of the global state-action pair $(s,a)$, thereby incurring the centralized information collection process. 
\textcolor{black}{A network cloud can be employed to train the actor NNs and critic NN jointly. By doing so, the SGD updates in \eqref{eq:actor_update} and \eqref{eq:critic_update} can be realized with the aid of the shared reward function \eqref{eq:reward}. 
Notice that the critic NN is employed only in the training phase for the optimization of the actor NNs. Therefore, once the training is completed, the critic NN can be discarded, and only the optimized actor NNs are dispatched to their desired agents for real-time task offloading and positioning decisions.} 

Since the proposed C-MADDPG framework facilitates individual computation architectures of the actor NNs, the decentralized execution of the UAV and ID agents is guaranteed based on Algorithm \ref{Algorithm 1}. \textcolor{black}{In the decentralized inference step, the agents share the coordination messages generated from the trained message actor NNs, whereas no parameter exchanges are required for the parameter sharing policy. In fact, this can be easily ensured in the training step since they are optimized under the identical GD update rule \eqref{eq:actor_update}. This offline training process incurs no additional communication overheads in the online inference step.} 

\begin{table}[t]
    \caption{\textcolor{black}{Simulation Parameters}}
    \label{table:Parameters}
    \centering
\begin{tabular}{|c|c||c|c|}
\hline
Symbol & Settings & Symbol & Settings \\
\hline
$\tau, T$ & $0.2\ \text{s}, 10$ & $\chi_{\text{LoS}},\chi_{\text{NLoS}}$ & $3\ \text{dB}, 23\ \text{dB}$\\
$I_{j}, \delta$ & $[2, 20]\ \text{Gbits}, 0.2$ & $K_{1}, K_{2}$ & $11.95, 0.14$\\
$C, \vartheta$ & $1550, 10^{-28}$ & $B, N_{0}$ & $10\ \text{MHz}, -130\ \text{dBm}$ \\
$\rho_{0}, \alpha$ & $-38 \ \text{dB}, 2$ & $\alpha_{\text{C}},\alpha_{\text{A}}$ & $1\times 10^{-3},\ 1\times 10^{-4}$\\
$p_{U}, p_{D}$ & $1\ \text{W}, 10\ \text{W}$& 
$v_{\max}$, $f_{\max}$ & $50$ m/s, $40$ GHz  \\
$\rho$, ${\sigma}^2$ & $0.45, 0.9995$& & \\
\hline
\end{tabular}
\end{table}

\section{Numerical Results}
\label{sec:simulation}
We present numerical results validating the proposed C-MADDPG. Unless otherwise stated, simulation parameters are fixed as in Table \ref{table:Parameters}. The critic NN has four fully-connected hidden layers each with $512$, $256$, $128$, and $64$ neurons. The message actor NNs $\mu_{U}(\cdot;\varphi_{U})$ and $\mu_{I}(\cdot;\varphi_{I})$ are built with three layers each having $128$ neurons. Also, we leverage four fully-connected layers with $128$ neurons for constructing the solution actor NNs $\pi_{U}(\cdot;\theta_{U})$ and $\pi_{I}(\cdot;\theta_{I})$. Output layers of the message actor NNs adopt the ReLU activation functions, whereas those of the solution actor NNs are set to the hyperbolic tangent function. Total $E=10^{5}$ episodes and $T=10$ time slots are considered in the training where each time slot consists of $|\mathcal{B}|=256$ mini-batch samples. For the initialization, the UAV and IDs are uniformly distributed in a $100$ m-by-$100$ m square area, and the altitude of the UAV is restricted to a bounded range $[0\ \text{m}, 60\ \text{m}]$. The trained C-MADDPG is tested with $10^{4}$ episodes.

We consider the following benchmark schemes.
\begin{itemize}
  \item \textit{Vanilla MADDPG} \cite{RLowe_MADDPG:Neur2017}: Message exchanges among the agents are not allowed. Thus, each agent is equipped with the solution actor NN only and produces $x_{j}^{(t)}$ based on the partial observation $o_{j}^{(t)}$.
  \item \textit{Single agent DDPG (SADDPG)}: An ideal centralized DRL scheme is adopted where a super actor NN decides the solution actions of all the UAV and ID agents jointly based on the state input $s^{(t)}$.
  \item \textit{C-MADDPG with GraphSage (C-MADDPG-GS)}: Instead of the GAT, the actor NN of the UAV agent is implemented with the GraphSage \cite{GCN:19}.
  \item \textit{Naive}: The UAV position is simply determined as a centroid of the IDs. Then, all IDs offload their tasks to the UAV, and the computing resources are equally allocated.
\end{itemize}
The vanilla MADDPG is a special case of the proposed C-MADDPG with no messages exchanged among agents. The SADDPG assumes an ideal centralized system where the UAV and IDs can share their observations perfectly. 
Unlike the proposed C-MADDPG, a single actor NN architecture of the SADDPG fails to achieve the scalability for varying $N$. Therefore, the SADDPG needs to be trained at each given $N$. For this reason, the SADDPG baseline provides unachievable upperbound performance of the proposed C-MADDPG approach. In the C-MADDPG-GS, the intermediate vector $\mathbf{w}_{j}^{(t)}$ of agent $j$ in \eqref{eq:DW_Mess_gene} is computed as the concatenation of the corresponding feature vector $\mathbf{e}_{j}^{(t)}$ in \eqref{eq:UAV_fea_ext} and the aggregation of others $\mathbf{e}_{k}^{(t)} (\forall k\in\tilde{\mathcal{N}}\backslash{\{j\}}$)~as
  \begin{align}
      \mathbf{w}_{j}^{(t)}=\left\{ \mathbf{e}_{j}^{(t)}, \sum_{k\in\tilde{\mathcal{N}}\setminus \{j\}}\mathbf{e}_{k}^{(t)} \right\}.
\end{align} 
In the C-MADDPG-GS, we halve the dimension of the feature vector $\mathbf{e}_{j}^{(t)}$ so that the concatenation $\mathbf{w}_{j}^{(t)}$ occupies the identical RBs with the proposed C-MADDPG.

\begin{figure}
\includegraphics[width=\linewidth]{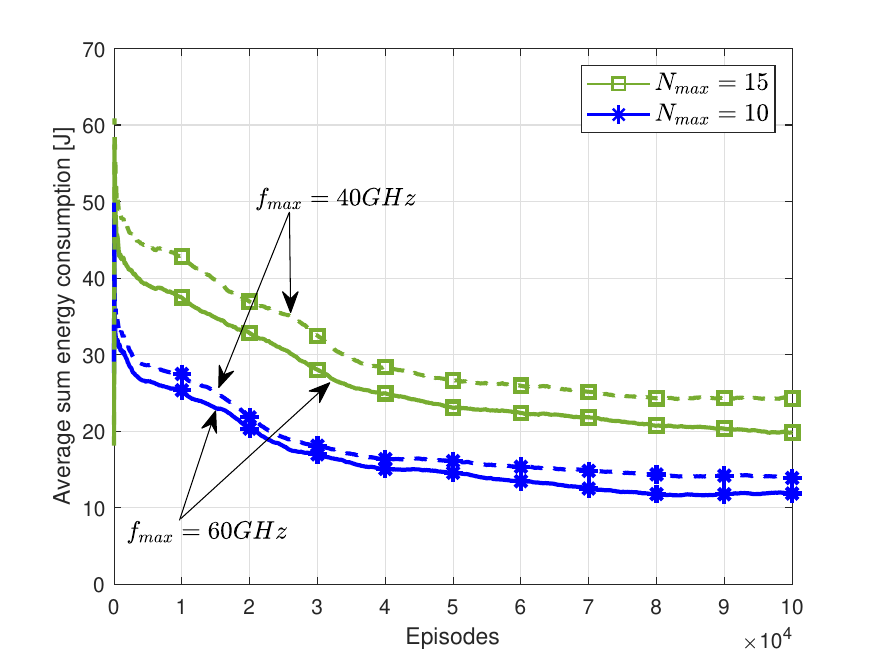}
\centering
\caption{\textcolor{black}{Convergence of the training process for different $N_{\max}$}}
\label{figure:num_of_episodes}
\end{figure}

Fig. \ref{figure:num_of_episodes} depicts the convergence behavior of the training process of the proposed C-MADDPG method in terms of the total energy consumption performance. \textcolor{black}{We plot the moving average energy consumption over $10^{4}$ episodes.} From the figure, we can check that the proposed C-MADDPG converges within $10^5$ episodes for all simulated $N_{\max}$. This implies the effectiveness of the proposed training strategy for handling a number of IDs. Since a small $N_{\max}$ results in simple training processes, it is beneficial to set $N_{\max}=10$. 

\begin{figure}
\includegraphics[width=\linewidth]{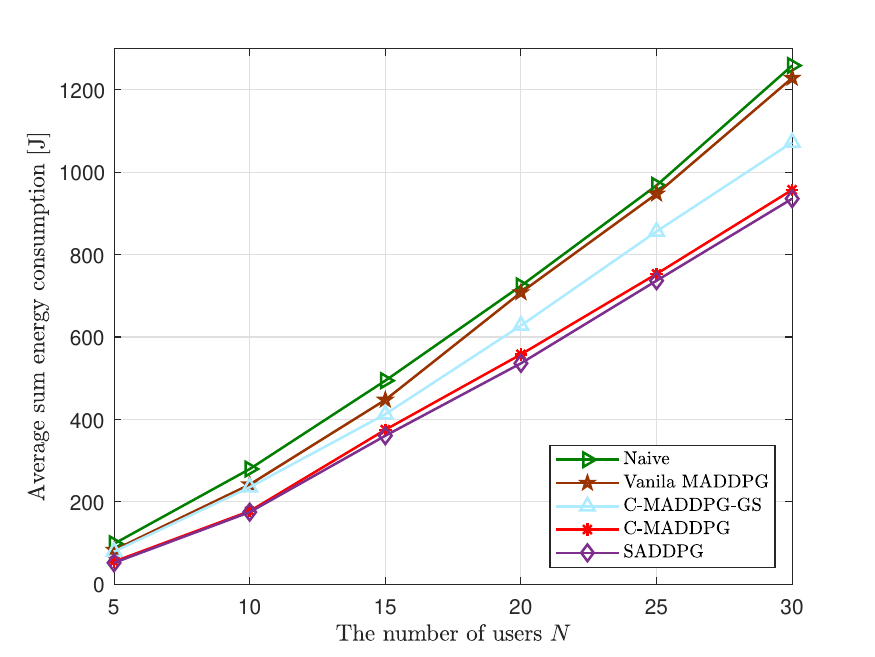}
\centering
\caption{Average sum energy consumption with respect to $N$}
\label{figure:num_of_users}
\end{figure}

\textcolor{black}{Fig. \ref{figure:num_of_users} compares the energy consumption performance of various schemes by changing $N$.} We set the ID population regime in the training to $N\in[5,10]$, and the test performance of the trained C-MADDPG is evaluated over $N\in[5,30]$. Thus, the actor NNs cannot observe training samples with $N\geq10$. Nevertheless, the proposed C-MADDPG shows negligible loss to the upperbound performance generated by the ideal centralized SADDPG method which is trained at each given $N$. This proves the scalability of the proposed approach where the actor NNs optimized at a small $N$ can be readily applied to larger networks. The gap between the vanilla MADDPG and the naive benchmark scheme decreases as $N$ grows. Without a proper agent interaction mechanism, the solution actor NNs of the vanilla MADDPG cannot provide efficient MEC management solutions and they simply converge to a suboptimal policy of the naive baseline. Based on this result, we can conclude that the message actor NNs play crucial roles in controlling the UAV and IDs in a decentralized manner. Also, the performance of the C-MADDPG-GS is degraded compared to the ideal SADDPG method. A simple sum pooling operation of the GS method fails to capture the importance of each ID agent in generating the downlink coordination message at the UAV agent. For this reason, the C-MADDPG-GS presents a large performance gap to the C-MADDPG. Thus, it is concluded that the GAT-based UAV message actor design is essential to control separate ID agents in a decentralized manner. 

\begin{figure}
\includegraphics[width=\linewidth]{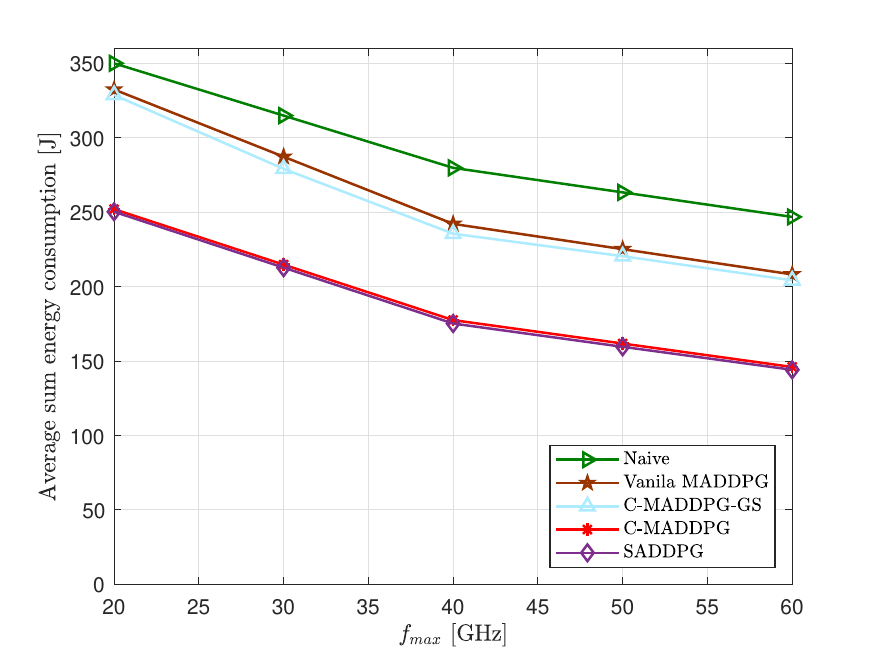}
\centering
\caption{Average sum energy consumption with respect to $f_{max}$}
\label{figure:f_max}
\end{figure}

Fig. \ref{figure:f_max} exhibits the energy consumption performance of various schemes by changing the maximum CPU frequency $f_{\max}$ with $N=10$. The UAV-aided MEC network can save the operating energy of the IDs as $f_{\max}$ increases. As expected, the proposed C-MADDPG outperforms other baseline schemes and provides almost identical performance to the upperbound SADDPG method. This validates the effectiveness of the C-MADDPG for optimizing appropriate MEC management solutions regardless of $f_{\max}$.

\begin{figure}
\includegraphics[width=\linewidth]{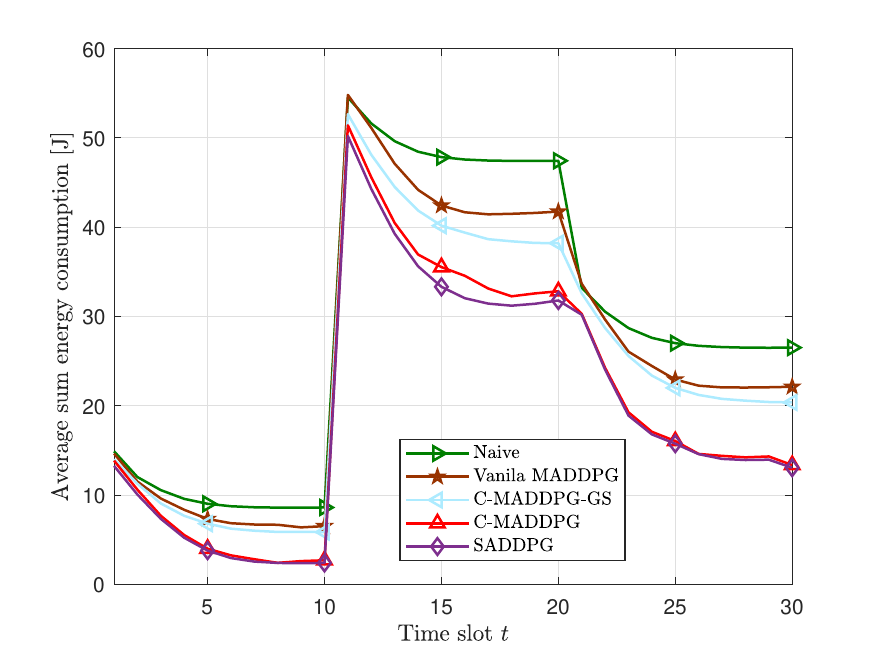}
\centering
\caption{Average sum energy consumption with varying $N$}
\label{figure:time_slot}
\end{figure}

The adaptability to time-varying ID population $N$ is investigated in Fig. \ref{figure:time_slot} which depicts the energy consumption as a function of the time slot. The number of IDs is assumed to change at every 10 time slots. We set $N=5$ for the first 10 time slots and then changes to $N=15$ for the next 10 time slots. For the last 10 time slots, the ID population is fixed as $N=10$. The energy consumption of all schemes highly fluctuates at the transition time slots and then is gradually reduced as the actor NNs yield convergent policies. Since the vanilla MADDPG relies on long-term collaboration through rewards rather than real-time cooperation based on observations, it adapts slowly to environment changes. In contrast, the proposed C-MADDPG shows fast convergence by means of online information exchange among agents. Thanks to the scalability, the proposed C-MADDPG can handle such highly fluctuating MEC configurations only with a sole training process. On the contrary, the non-scalable baselines such as the SADDPG and vanilla MADDPG resort to several trained actor NNs dedicated to each possible $N$. Also, the proposed approach with the GAT architecture achieves the performance of the ideal SADDPG method, proving the effectiveness of the proposed scheme.

\begin{table}[t]
   \caption{\textcolor{black}{Average CPU running time [msec]}}
   \label{table:CPU_time}
   \centering
\begin{tabular}{|c||c|c|c|c|c|}
\hline
$N$ & 10 & 15 & 20 & 25 & 30 \\
\hhline{|=#=|=|=|=|=|}
SADDPG & 2.30 & 2.33 & 2.35 & 2.37 & 2.39  \\ \hline
C-MADDPG & 2.19 & 2.19 & 2.19 & 2.19 & 2.19 \\ 
\hline
\end{tabular}
\end{table}

\textcolor{black}{Table \ref{table:CPU_time} compares the inference time complexity of trained actor NNs by evaluating the average CPU running time for executing $10^{4}$ episodes. The computation complexity of the proposed C-MADDPG remains unchanged with ID populations due to its decentralized and parallel architecture. For this reason, the inference complexity of the proposed C-MADDPG is slightly lower than that of the centralized SADDPG without incurring performance degradation.}

\section{Conclusion}
\label{sec:conclusions}
\textcolor{black}{This work has proposed a novel C-MADDPG approach for the decentralized control of the UAV-aided MEC networks where the UAV and IDs can only get access to their local observations only. 
To build a valid decentralized decision-making policy, it is necessary to develop a proper coordination protocol, in particular, interaction messages bearing sufficient statistics of the optimal solutions of others. The considered problem has been formalized into a cooperative multi-agent POMDP which includes interaction messages as well as solutions of individual agents as action variables. Such dual actions request two different actor NNs, the message actor NN and solution actor NN, each of which accounts for the agent coordination and solution optimization. For effective message aggregation, the message actor NN at the UAV adopts the GAT architecture. Along with the parameter sharing policy, this graph-inspired structure leads to versatile operations that do not depend on the network size. 
Also, the joint training algorithm of all actor NNs has been proposed. To achieve the scalability to the ID population, the proposed C-MADDPG has been optimized over arbitrary random number of the ID agents. Numerical results have demonstrated the superiority of the proposed scheme over existing schemes.}
\bibliographystyle{IEEEtran}
\begingroup
\input{bibliography.filelist}
\endgroup
\end{document}